\documentstyle[preprint,aps]{revtex}
\begin{document}
\bibliographystyle{prsty}
\draft
\title{Experimental realization of a highly structured search algorithm}
\author{Xiwen Zhu, Ximing Fang, Mang Feng, Fei Du, Kelin Gao and
Xi'an Mao} 
\address{Laboratory of Magnetic Resonance and Atomic and Molecular Physics,\\
Wuhan Institute of Physics and Mathematics, Chinese Academy of
Sciences,\\ Wuhan 430071, P. R. China}
\date{\today}
\maketitle
\begin{abstract}
The highly structured search algorithm proposed by Hogg[Phys.Rev.Lett. 80,2473(1998)]
is implemented experimentally for the 1-SAT problem in a single search step by 
using nuclear magnetic resonance technique with two-qubit sample. It is the
first demonstration of the Hogg's algorithm, and can be readily
extended to solving 1-SAT problem for more qubits in one step  if the
appropriate samples possessing more qubits are experimentally feasible.
\end{abstract}
\vskip 0.5cm

\narrowtext

Quantum computers[1-5] can outperform classical ones owing to
utilization of quantum mechanical effects. The potential
for quantum parallel computing mainly resulted from superpositions and
entanglements of quantum states has made quantum computers capable of solving 
classically intractable problems (such as factoring
large integers[3]) and finding tractable solutions more rapidly(e. g.,
searching an unsorted database[6]).  However,  carrying  out  an 
actual quantum 
computing is much difficult although the rudiment of the quantum computing is 
easily understood and several quantum algorithm have been discovered.
Among the approaches meeting the requirement for performing quantum computing,
the nuclear magnetic resonance(NMR) technique[7,8] is considered to be
one of current promising method, with which the Deutsch-Jozsa algorithm[9]
and Grover's search one[6] has been experimentally implemented  with
two-qubit samples[10-13]. Recently, it was reported that an algorithmic
benchmark has been experimentally realized using NMR with seven qubits[14].
It was easily found from those works that NMR can
act as a considerably satisfactory small-scaled quantum computer to
test the simple cases of the various quantum algorithm. In this contribution, 
also using NMR scheme with two-qubit sample, we report the first 
experimental demonstration of another quantum search algorithm. This algorithm, called a highly 
structured 
quantum search algorithm[15], was claimed to be much more efficient than Grover's 
algorithm[6] that ignores the problem structure, and can be exploited in many 
practical examples instead of for academic study alone as an artificial 
problem[9].

The Hogg's algorithm is associated with the combinatorial searching known as
nondeterministic polynomial-time problems(NP)[16], an important class of
intractable problems. Its prototype is the satisfiability problems(SAT)[16].
A SAT consists of a logic formula in $n$ boolean variables, V$_{1}$,...,V$_{n}$, 
and the requirement to find an assignment, specifying a value for each binary
variable, that makes the formula true. The logic formula can be expressed as
a conjunction of $m$ clauses and each clauses is a disjunction of some
variables. When all the clauses have exactly $k$ variables, the problem is
called k-SAT, with 1-SAT being the simplest case.  Combinatorial search aims 
at seeking the solution of a SAT from
the total of 2$^{n}$ assignments. In general, the computational cost of
solving a SAT grows exponentially with $n$ in the worst case, making the SAT 
one of the most difficult NP problems[16]. For a few simple cases like 1-SAT 
and 2-SAT, however,
solutions can be quickly and accurately determined without exponential cost
using the regular structure of the problem. As each false clause for a
given assignment is counted as a conflict, solutions are assignments with no
conflicts. The classical search cost
for such cases is O($n$) due to each clause eliminating one value for a single 
variable, but the quantum counterpart with Hogg's algorithm for 1-SAT
and maximally constrained k-SAT a single step[15],
with the help of quantum parallelism and interference combined with the 
problem structure, which shows very high efficiency for large $n$.

Actually, Hogg's algorithm consists of three stages[15]. First, one makes
an equal superposition of bases, $|s>$, of the $n$ two-state quantum system as 
the initial state $|\psi _{i}>=2^{-\frac{n}{2}}\sum_{s}|s>$, with bit strings 
$s$ being all 2$^{n}$
assignments of $n$ variables. Secondly, a transformation R is applied on 
$|\psi _{i}>$. The elements R$_{ss}$ of the diagonal matrix R are determined
by the conflicts c$_{s}$ of $s$, which are governed by the constraints
expressed in clauses of the logic formula.It is explicitly expressed as
\[ R(s)=\left\{
\begin{array}{ll}
\sqrt {2} \cos{(2c-1)\frac {\pi}{4}}&~~~~{\rm for~even~m}\\
i^{c}&~~~~{\rm for~odd~m .~~~~~~~~~~~~~~~~~~~~~~~~~(1)}
\end{array} \right. \] 
Finally, an operation U with the form
\[U_{rs}=\left\{
\begin{array}{ll}
2^{-\frac {n-1}{2}} \cos {[(n-m+1-2d)\frac{\pi}{4}]} &~~~~ {\rm for~even~m}\\
2^{-\frac{n}{2}} e^{i (n-m) \frac{\pi}{4}}(-i)^{d} &~~~~ {\rm for~odd~m} ,
\end{array} \right.\]
is acted. It can be seen that the matrix elements U$_{rs}$ depend only on the 
Hamming distance d$_{r,s}$ between $r$ and $s$. As it describes the inconsistency of
all the corresponding bits of bit strings $r$ and $s$, d$_{r,s}$ has some
correlation with conflicts c$_{s}$. By exploiting the intrinsic relation
between d$_{r,s}$ and c$_{s}$ and ingeniously constructing R$_{ss}$ and 
U$_{rs}$with d$_{r,s}$ and c$_{s}$, making the combined transformation UR on 
$|\psi _{i}>$ once is able to find the expected state, i. e., UR $|\psi
_{i}>= $ $|\psi _{0}>$, which has equal amplitudes among solutions and no
amplitudes among nonsolutions. 

In the view of quantum physics, Hogg's algorithm subtly makes use of
the properties of quantum superposition and interference. When the calculating 
system is prepared on the equal superposition state,
$|\psi_{i}>=2^{-\frac{n}{2}} \sum_{s}|s>$, all the possible values,
which can be assigned to the logical formula, regardless correct or
incorrect, are embodied in the state.Then the state is subjected to the
transform $R$, which gives a varied phase to each component state of the
superposition  state according to its conflicts $c_{s}$. The final operation
$U$ destructively annihilates every component state that its conflict is not 
zero, construtively enhances the correct component state that its conflicts is zero, finds
out the exact component state that satisfies the logical formula. In
one word, the essence of searching achieved in a single
step is to examine all the assignments simultaneously and to make use of
quantum interference embedded in UR. By a single step we mean the number of
search steps or sequentially examined assignments, not the number of
elementary computational operations required.

Our implementation of Hogg's algorithm was performed using solution NMR,
applied to an ensemble of carbon-13 labelled chloroform molecules $(^{13}$%
CHCl$_{3})$. Two heteronuclear spins $^{1}$H and $^{13}$C in $^{13}$CHCl$%
_{3} $ work as two qubits in our quantum computing. Similar to other
experiments[9-12,14,17-19] of quantum information processing using NMR, all
the logic operations were realized by a specified sequence of
radio-frequency (RF) pulses and the spin-spin coupling between the nuclei,
and should be finished in the period of time much shorter than the
relaxation times T$_{1}$ and T$_{2}$ of the nuclei in order to minimize the
decoherence effect.

A quantum circuit for achieving this highly structured search in a two-qubit
system is shown in Fig. 1. The algorithm starts in the state $|\psi
_{s}>=|00>$, labelling the states of $^{1}$H and $^{13}$C spins from left to
right respectively. As the state of a bulk sample at room temperature is
described by a density matrix $\rho _{th}$ for a thermally equilibrated
system,we prepared an effective pure state $\rho =|00><00|$ by temporal 
averaging[20],
with which a deviation density matrix with diagonal elements of diag($\rho
_{s}$)=[1,0,0,0] could be extracted from the summation over cyclic
permutations of the populations of the $|01>$, $|10>$, and $|11>$
states three times on the thermal matrix $\rho _{th}$. The Hadamard gates H
in Fig. 1 were implemented by using RF pulse sequence $\left( \frac{\pi }{2}%
\right) _{y}\left( \pi \right) _{x}$ or its equivalents and thus a uniform
superposition state denoted by the density matrix $\rho _{i}=|\psi
_{i}><\psi _{i}|$ with diag($\rho _{i})$=[1,1,1,1] was obtained. Then two
crucial operations R and U were applied, which are related to the problem
structure and expressions for the logic formula and lead to the enhancement
of solution states and the cancel of nonsolution states.

For our two-qubit system(i. e., $n=2$), $m$, the number of clauses in the
1-SAT formula with satisfiable solutions can be assumed 2 or 1. When
$m=3$ and 4, no satisfying assignment exists due to over-constrained
formula being unsatisfiable. Four satisfiable formulas in the $m=2$
case are $V_{1}\wedge V_{2}$, $V_{1}\wedge \overline{V}_{2}$, $\overline{V}%
_{1}\wedge V_{2}$ and$\overline{V}_{1}\wedge \overline{V}_{2}$, where $%
\overline{V}_{i}$ ($i=1,2$) are the negation of $V_{i}$. The corresponding
solutions are easily found to be $|11>$, $|10>$, $|01>$ and 
$|00>$ respectively, where the logic variable
with subscript 1 corresponds to the high qubit and 2 the low qubit. Similarly,
the formulas for $m$=1 are $V_{2}$, $V_{1}$, $\overline{V}_{2}$ and $%
\overline{V}_{1}$, corresponding to the unnormalized answer states 
$|01>+|11>$, $|10>+|11>$, $|00>+|10>$ and $|00>+|01>$, respectively. According 
to the number
of states in the solutions, they can be called as single- and multiple- item
searching. The algorithm on two instances of 1-SAT was experimentally 
demonstrated: one for $m=2$ with clauses $V_{1}$ and $V_{2}$, and the
other for $m=1$ with clause $V_{2}$.  

First of all, we have to derived the expressions for R and U from Ref.[15]. 
With Eq.(1), after calculating the conflicts $c$ of each
string assigned to the logical formula respectively, the elements of 
the diagonal matrices 
R$_{V_{1}\wedge V_{2}}$ and R$_{V_{2}}$ were calculated to be $%
diag(R_{V_{1}\wedge V_{2}})=[-1,1,1,1]$ and $diag(R_{V_{2}})=[i,1,i,1]$. U
could be represented by $W\Gamma W$, where $W=H_{1}\otimes H_{2}$ with $%
H_{i} $ being the Hadamard gates, and $\Gamma$ is a diagonal matrix
with the diagonal elements 
\[ \Gamma_{rr}=\left\{
\begin{array}{ll}
\sqrt {2} \cos{[(m-2|r|-1)\frac {\pi}{4}}&~~~~{\rm for~even~m}\\
i^{|r|}e^{-im\frac{\pi}{4}}&~~~~{\rm for~odd~m} .
\end{array} \right. \] 
Elements of the diagonal matrices $\Gamma$ for $m=2$ and $m=1$ were evaluated 
respectively as $diag(\Gamma _{m=2})=[1,1,1,-1]
$ and $diag(\Gamma _{m=1})=[1,i,i,-1]$. 

In order to realize the transformations with NMR, we designed the following RF
pulse sequences respectively:

$R_{V_{1}\wedge V_{2}}$: $\left( \frac{\pi }{2}\right) _{y_{1}}\left(
\frac{\pi }{2}\right)
_{x_{1}}\left( \frac{\pi }{2}\right) _{-y_{1}}-\frac{1}{2J}-\left( \frac{\pi 
}{2}\right) _{y_{2}}\left( \frac{\pi }{2}\right) _{x_{2}}\left( \frac{\pi }
{2}\right) _{-y_{2}}$

$R_{V_{2}}$: $\left( \frac{\pi }{2}\right)
_{y_{1}}\left( \frac{\pi }{2}\right) _{x_{1}}\left( \frac{\pi }{2}\right)
_{-y_{1}}$ 

$\Gamma _{m=2}$:$\left( \frac{\pi }{2}\right) _{y_{1}}\left( 
\frac{\pi }{2}\right) _{-x_{1}}\left( \frac{\pi }{2}\right) _{-y_{1}}-
\frac{1}{2J}-\left( \frac{\pi }{2}\right) _{y_{2}}\left( \frac{\pi }{2}\right)
_{-x_{2}}\left( \frac{\pi }{2}\right)_{-y_{2}}$ %\linebreak %

$\Gamma _{m=1}$: 
$\left( \frac{\pi }{2}\right) _{y_{1}}\left( \frac{\pi }{2}\right)
_{x_{1}}\left( \frac{\pi }{2}\right) _{y_{1}}\left( \frac{\pi }{2}\right)
_{y_{2}}\left( \frac{\pi }{2}\right) _{-x_{2}}\left( \frac{\pi }{2}\right)
_{-y_{2}}$ 

%Although the pulse sequences of every single transform were designed,
%the whole quantum calculation can not be completed well by simply applying
%them one after anther because of the length of coherence time and the error
%accumulation of the pulses. In order 
To make the best use of the available coherence time
and to diminish errors due to the increased number of RF pulses, we finally
optimized or reduced the pulse sequence constructed from the executing
pulses of H, $\Gamma $ and R mentioned above with the help of NMR
principle[21]. The reduced pulse sequence for (UR)$_{V_{1}\wedge V_{2}}$ and
(UR)$_{V_{2}}$ shown in Fig. 2 were applied upon $\rho _{i}$, resulting the
output matrix $\rho _{0}\equiv |\psi _{0}><\psi _{0}|=(UR)\rho
_{i}(UR)^{\dagger }$.

In order to read out the
results of Hogg's algorithm accurately, a method of quantum state
tomography[22] was adopted to reconstruct all the elements of the output
density matrix $\rho _{0}$. For this end, we applied different read-out
pulses immediately after the searching pulses UR, one for each run.
Explicitly, nine pulses E$_{1}$E$_{2}$, E$_{1}\left( \frac{\pi }{2}\right)
_{x_{2}}$, E$_{1}\left( \frac{\pi }{2}\right) _{y_{2}}$, $\left( \frac{\pi }{%
2}\right) _{x_{1}}$E$_{2}$, $\left( \frac{\pi }{2}\right) _{x_{1}}
\left(\frac{\pi }{2}\right) _{x_{2}}$, $\left( \frac{\pi }{2}\right)
_{x_{1}}\left( \frac{\pi }{2}\right) _{y_{2}}$, 
$\left(\frac{\pi }{2}\right)_{y_{1}}$E$_{2}$, 
$\left(\frac{\pi }{2}\right)_{y_{1}}\left( \frac{\pi }{2}\right)
_{x_{2}}$ and $\left( \frac{\pi }{2}\right) _{y_{1}}\left( 
\frac{\pi }{2}\right) _{y_{2}}$were exploited, where E$_{i}$ $(i=1,2)$
means no pulse acted on the i-th spin. Then we calculated the signal
intensities by integrating the proton and carbon spectral lines acquired in
each run and reproduced the whole matrix $\rho _{0}$ using the least-square
fitting to the coupled equations connecting the signal intensities with the
matrix elements. The recovered matrices $\rho _{0}$ from the experimental
data were depicted in Fig. 3, along with the theoretical ones for
comparison. It is clearly seen from Fig. 3 that only the elements of 
$<11|\rho _{0}(V_{1}\bigwedge V_{2})|11>$, $<01|\rho_{0}(V_{2})|01>$, 
$<01|\rho _{0}(V_{2})|11>$, $<11|\rho _{0}(V_{2})|01>$ and 
$<11|\rho _{0}(V_{2})|11>$ are significant, which justifies the searched 
items to be $|11>$ for $V_{1}\bigwedge V_{2}$ and unnormalized $|01>+|11>$ 
for $V_{2}$ formula. Due to experimental errors, however, several elements that
should be zero theoretically still have small amounts of modulus, with the
maxima being 8\% for single-item and 19\% for multiple-item searching.

Errors in the experiments result from several sources. The effective pure
state $|00>$ prepared by temporal averaging is
generally not ideal, and small residual populations in the states 
$|01>$, $|10>$ and $|11>$ of the deviation density matrix will
surely cause errors in the later stage of experiments. The inhomogeneity of
RF fields and static magnetic fields and imperfections of the pulse-length
calibration are also important sources of errors. Although we have tried to
minimize these effects by careful adjustments of and measurements with the
apparatus, some factors were still not well under control. Lastly but not
insignificantly, losses of the coherence and populations of the density
matrices in the whole process of experiments will lead to deviations of
measured data from those expected by theory, especially for the off-diagonal
elements. Perhaps that is why the maximum error in the multiple-item
searching is larger than that in the single-item case. In all senses,
minimizing experimental errors in all available ways is essential to
accurately operating NMR quantum computers.

We have experimentally demonstrated a quantum algorithm for the highly
structured combinatorial searching and found solutions to the 1-SAT problem
in one algorithmic step almost surely using an NMR quantum computer with 2 qubits. By 
contrasting with other search methods that ignore the problem structure 
requiring O(2$^{n}$) steps classically and O(2$^{\frac{n}{2}}$) steps on 
quantum computers[6], the searching with Hogg's algorithm can be accomplished
in principle more 
efficiently. However, there is no difference between Hogg's algorithm
and Grover's one when only two qubits are considered. The potential of
the high efficiency owned by Hogg's algorithm will be displayed in solving
various 1-SAT problems for $n > 2$ qubits in one step  with current NMR
technique, which can be readily extended from the present experiment, if the 
appropriate samples possessing more qubits can be experimentally 
feasible.
Furthermore, Hogg's algorithm for 1-SAT
could be generalized to the maximally constrained k-SAT problems[15], which
have practical examples such as scheduling, finding low energy states of
spin glasses and proteins, and automatic theorem proving. While scaling up
an NMR computer to much large systems poses daunting challenges, building
such a device with a few qubits by some creative approaches for
demonstrating algorithms stated above and the others is promising.

We thank Xijia Miao for help in the early stage of
experiments. This work was supported by the Chinese Academy of Sciences and
National Natural Science Foundation of China.

\newpage

\begin{center}{\bf Captions of the figures}\end{center}

Figure 1~~ A quantum circuit for implementing a highly structure search
algorithm on a two-qubit computer. Two Hadamard gates H$\otimes $H transform
an effective pure states $|\psi _{s}>=|00>$ into a uniform
superposition state $|\psi _{i}>$, which is then
transformed to the answer state $|\psi _{0}>$ after the
action of gates R and U. For the definition of R and U, see text.

Figure 2~~ The reduced NMR pulse sequences used to execute\ a) (UR)$_{V_{1}
\wedge V_{2}\,}$and b) (UR)$_{V_{2}}$\thinspace operations. Narrow and
wide boxes correspond to $\frac{\pi }{2}$ and $\pi $ pulses respectively. X
and Y denote the pulses along the x- and y-axis, $\overline{X}$ and$\;%
\overline{Y},$ opposite to the x- and y-axis. The time period $\tau $ is set
equal to $\frac{1}{4J}$, where J is the size of the spin-spin coupling
between nuclei $^{1}H$ and $^{13}C$. Experiments were performed using a
Bruker ARX500 spectrometer. $^{13}C$-labelled CHCl$_{3}$ were obtained from
Cambridge Isotope Laboratories.

Figure 3~~ Experimentally recovered and theoretically expected deviation
density matrices, $\rho _{0},$ after completion of the combinatorial search
with the aim of satisfying logic formulas of both V$_{1}\wedge V_{2}$ and 
V$_{2}$. 
The ordinate represents the modulus of matrix elements of $\rho _{0}$ (not
normalized). The numbers 0, 1, 2 and 3 in the horizontal plane denote the
subscripts $|00>$, $|01>$, $|10>$ and $|11>$ of the elements, respectively. 
a) Experiment for $V_{1}\wedge
V_{2}$, b) Theory for $V_{1}\wedge V_{2}$, c) Experiment for $V_{2}$ and d)
Theory for $V_{2}$.

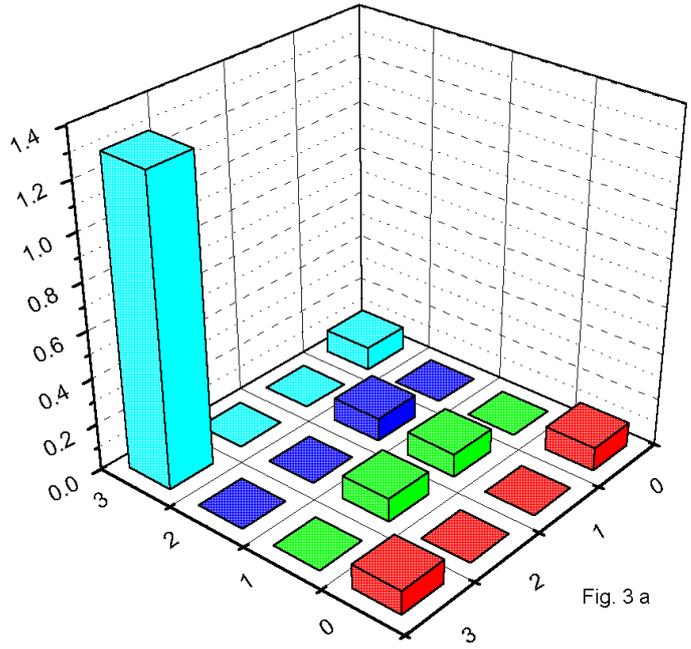
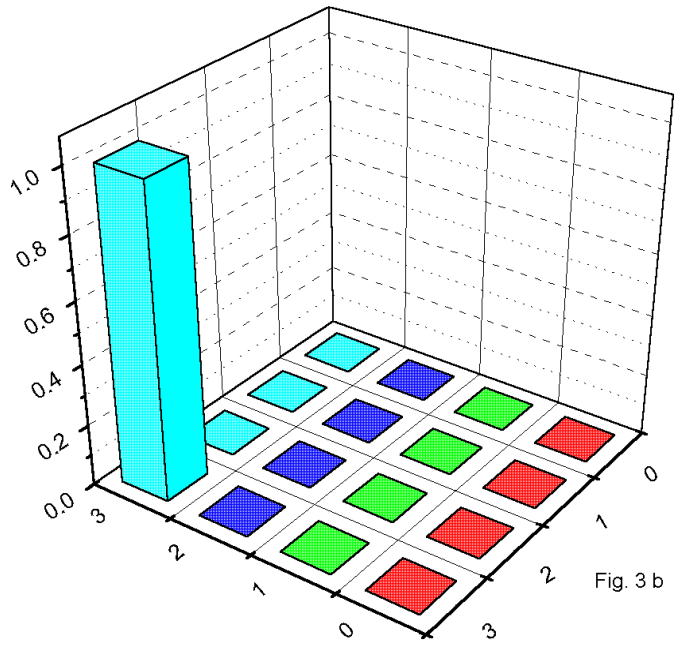
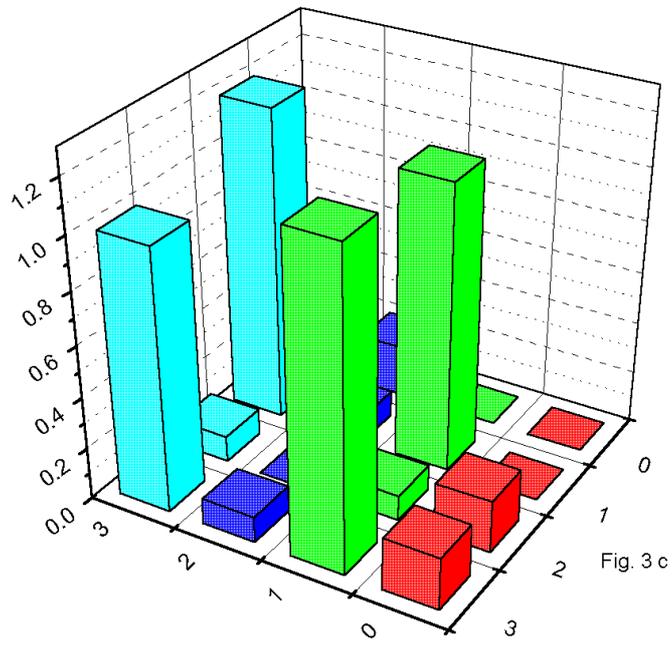
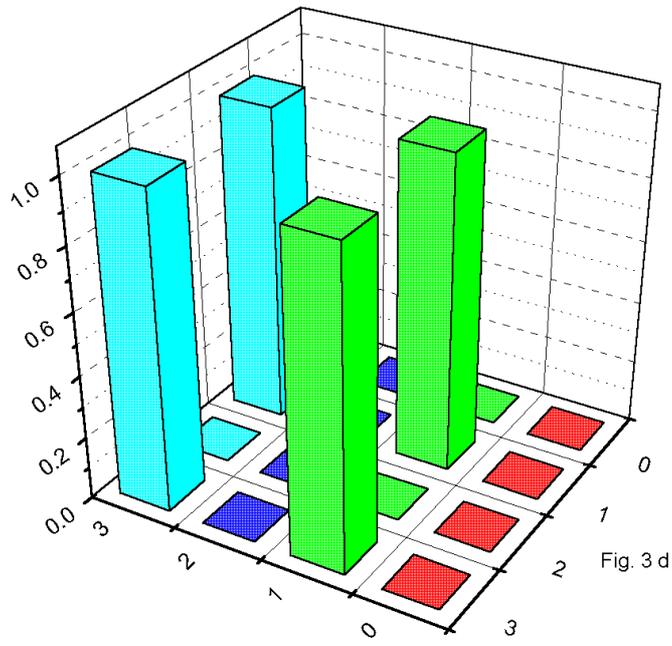
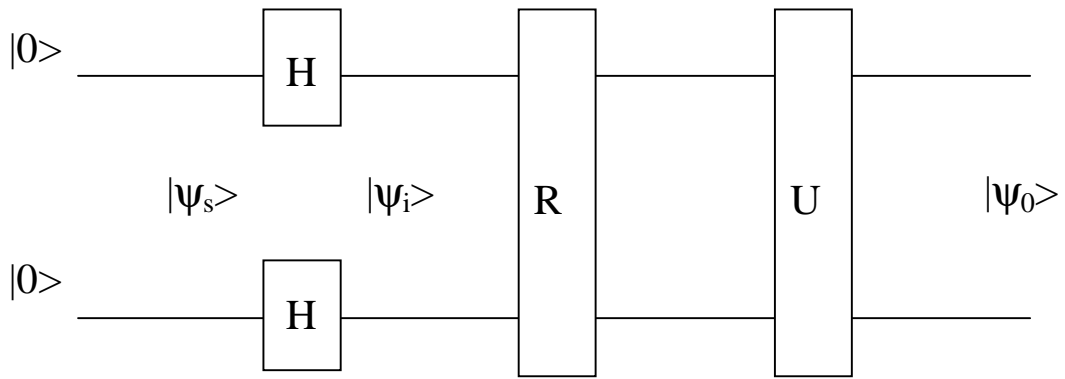
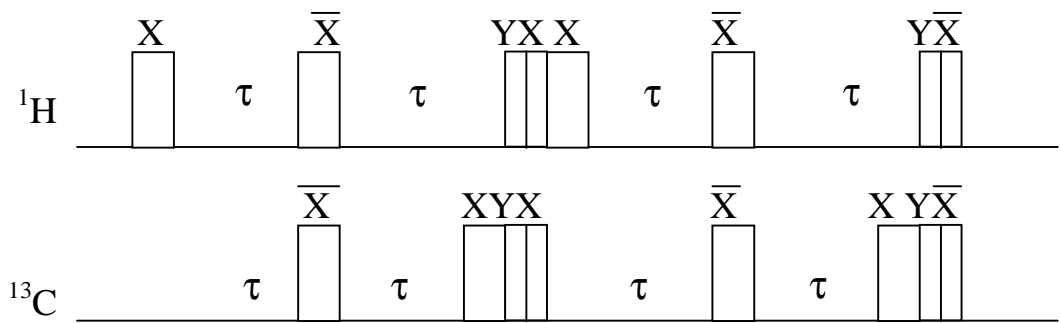
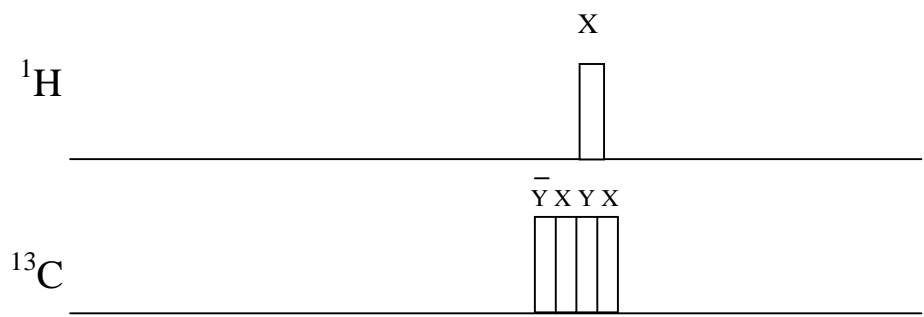
\end{document}